# Private Product Computation Using Quantum Entanglement

**RENÉ BØDKER CHRISTENSEN[1,2] AND PETAR POPOVSKI[2] (Fellow, IEEE)**

[1]Department of Mathematical Sciences, Aalborg University, 9220 Aalborg ø, Denmark
[2]Department of Electronic Systems, Aalborg University, 9220 Aalborg ø, Denmark

Corresponding author: René Bødker Christensen (e-mail: rene@math.aau.dk).

This work was supported by the Villum Investigator Grant "WATER" from the Velux Foundations, Denmark

**ABSTRACT** In this article, we show that a pair of entangled qubits can be used to compute a product privately. More precisely, two participants with a private input from a finite field can perform local operations on a shared, Bell-like quantum state, and when these qubits are later sent to a third participant, the third participant can determine the product of the inputs, but without learning more about the individual inputs. We give a concrete way to realize this product computation for arbitrary finite fields of prime order.

## I. INTRODUCTION

Having access to quantum bits, or *qubits*, opens up new methods that would be impossible in the classical realm. An example of this is superdense coding [1], where two participants sharing an entangled pair can send a single qubit to transmit two classical bits, see e.g., [2, pp. 97–98].

In this article, we explore how entangled pairs may be used to compute private products over finite fields. Namely, if two participants each hold an element of a finite field, and they wish to reveal the product of their elements to a third party without revealing their individual inputs, then this can in certain cases be achieved using pairs of entangled qubits. One may argue that it is also possible to achieve this using Shamir's secret sharing scheme [3], [4, Ch. 3]; this is true, in the same sense that one may similarly argue that superdense coding can be obviated by classical transmission of two bits using a single symbol, as in quadrature phase shift keying, rather than using entanglement. As such, our motivation for studying this problem is not so much the application to a specific real-world problem, but should be seen more as an exploration of the possibilities opened up by using quantum information processing.

The rest of this article is organized as follows. Section II describes our model assumptions and recalls the basic quantum properties that will be used throughout. After that, Section III provides a sketch of the problem in the case of products over $\mathbb{F}_2$. This is meant to provide a better intuition about the challenges and requirements in the general case. In Section IV, we then define properties necessary to compute a private product over general finite fields of prime size, and use this to formulate a general protocol. Section V provides an explicit construction of a private product family that can be applied in the general protocol, and in Section VI we show how this encoding can be realized systematically by Alice and Bob. Finally, Section VII indicates how the developed methods can be applied to (small) private set intersections (PSI) and to dot products, both in the binary case. Finally, Section VIII concludes this article and lists a few open problems for future research.

## II. PRELIMINARIES

Throughout this article, we assume that $p$ is a prime and let $\mathbb{F}_p$ denote the finite field of order $p$.

### A. MODEL ASSUMPTIONS

Keeping with cryptographic tradition, we will call the three participants Alice, Bob, and Charlie. Alice and Bob each hold an input $a \in \mathbb{F}_p$ and $b \in \mathbb{F}_p$, respectively, and their goal is for Charlie to learn $ab$. In addition, they need to achieve this in such a way that the following holds.

1) Alice does not learn anything about $b$.
2) Bob does not learn anything about $a$.
3) Charlie does not learn anything about $(a, b)$ except what is implied by $ab$.

We assume only a limited number of communication channels between the participants. Namely, we assume the existence of a classical channel[1] from Alice to Bob, and quantum channels from Alice to Charlie and from Bob to Charlie. For simplicity, we assume that all channels are perfectly private and error-free. Note that if the participants do not have access to other channels than the three above-mentioned, the classical solution provided by Shamir's secret sharing is no longer possible.

[1]As noted by one of the reviewers, one may replace this classical channel by shared randomness between Alice and Bob that is independent of $(a, b)$.

Throughout, we assume participants are "honest-but-curious." That is, they will follow protocols as specified, but they may try to use anything received during the protocol in an attempt to extract information about the other participants' inputs.

### B. QUANTUM ENTANGLEMENT

A $p$-ary quantum bit can be described by a state in $\mathbb{C}^p$, and we will fix an orthonormal basis $\{|i\rangle\}_{i\in\mathbb{F}_p}$ of $\mathbb{C}^p$. Here, $i$ does not refer to the imaginary unit, and throughout, we will simply use it as an index.

Now, let $\omega \in \mathbb{C}$ be a primitive $p$'th root of unity. As defined in [5], the $p$-ary bit flip and phase shift operators applied to the basis states are

$$X(a) : \begin{cases} \mathbb{C}^p \to \mathbb{C}^p \\ |i\rangle \mapsto |i+a\rangle \end{cases} \quad \text{and} \quad Z(b) : \begin{cases} \mathbb{C}^p \to \mathbb{C}^p \\ |i\rangle \mapsto \omega^{bi}|i\rangle \end{cases}$$

where $a, b \in \mathbb{F}_p$, and images of arbitrary states are defined by linearity of the operators.

Consider the Bell-like states given by

$$|\varphi_{ab}\rangle = \sum_{i=0}^{p-1} \omega^{bi}|i+a\rangle|i\rangle, \quad a, b \in \mathbb{F}_p. \tag{1}$$

With this definition, we have $|\varphi_{00}\rangle = \sum_{i=0}^{p-1}|i\rangle|i\rangle$, and $|\varphi_{ab}\rangle = (X(a) \otimes Z(b))|\varphi_{00}\rangle$. Note, that in the literature, $|\varphi^{\pm}\rangle$ and $|\psi^{\pm}\rangle$ are commonly used to denote the four Bell states in the binary case, but we use the notation in (1) as it eases notation in our setting.

The mathematical representation of a quantum state can be multiplied by a complex scalar of modulus 1, which is called a global phase. The significance of this global phase does not carry over to the physical qubit, however, as a global phase does not influence the outcomes when measuring a qubit [2]. For this reason, we will ignore global phase factors throughout most of this work.

## III. BINARY CASE

In order to illustrate the ideas in this work, we give a detailed overview in the case $p = 2$. Here, $\omega = -1$, and the states in (1) are given by

$$\begin{aligned} |\varphi_{00}\rangle &= \frac{|00\rangle+|11\rangle}{\sqrt{2}} & |\varphi_{01}\rangle &= \frac{|00\rangle-|11\rangle}{\sqrt{2}} \\ |\varphi_{10}\rangle &= \frac{|10\rangle+|01\rangle}{\sqrt{2}} & |\varphi_{11}\rangle &= \frac{|10\rangle-|01\rangle}{\sqrt{2}} \end{aligned} \tag{2}$$

where we use the notation $|ii\rangle = |i\rangle|i\rangle$.

Assume that Alice and Bob have already prepared the Bell state $|\varphi_{00}\rangle$ and split the qubits between them such that Alice holds the first qubit and Bob holds the second.

**TABLE 1. Three Encodings of $(a, b)$ for $\mathbb{F}_2$**

| | 0 | 1 |
|---|---|---|
| 0 | $I \otimes I$ | $I \otimes Z$ |
| 1 | $X \otimes I$ | $X \otimes Z$ |

| | 0 | 1 |
|---|---|---|
| 0 | $I \otimes Z$ | $I \otimes X$ |
| 1 | $Z \otimes Z$ | $Z \otimes X$ |

| | 0 | 1 |
|---|---|---|
| 0 | $X \otimes I$ | $X \otimes X$ |
| 1 | $Z \otimes I$ | $Z \otimes X$ |

**TABLE 2. Resulting States of the Encodings in Table 1**

| | 0 | 1 |
|---|---|---|
| 0 | $\|\varphi_{00}\rangle$ | $\|\varphi_{01}\rangle$ |
| 1 | $\|\varphi_{10}\rangle$ | $\|\varphi_{11}\rangle$ |

| | 0 | 1 |
|---|---|---|
| 0 | $\|\varphi_{01}\rangle$ | $\|\varphi_{10}\rangle$ |
| 1 | $\|\varphi_{00}\rangle$ | $\|\varphi_{11}\rangle$ |

| | 0 | 1 |
|---|---|---|
| 0 | $\|\varphi_{10}\rangle$ | $\|\varphi_{00}\rangle$ |
| 1 | $\|\varphi_{01}\rangle$ | $\|\varphi_{11}\rangle$ |

Alice and Bob now do the following. If $a = 1$, Alice will apply $X$ to her qubit, and if $b = 1$, Bob will apply $Z$. The reader may check that this maps $|\varphi_{00}\rangle$ to $|\varphi_{ab}\rangle$. If Alice and Bob send their individual qubits to Charlie, he can measure the received system in the Bell basis to recover $|\varphi_{ab}\rangle$. The problem with this approach, however, is that Charlie not only learns the product $ab$. He also learns the individual inputs $a$ and $b$ since $|\varphi_{ab}\rangle$ is the output if and only if Alice has input $a$ and Bob has input $b$.

In order to fix this, Alice and Bob will choose uniformly at random between three different encodings that all encode $(a, b) = (1, 1)$ to the same state $|\varphi_{11}\rangle$ (up to a global phase factor). That is, they choose one of the encodings in Table 1 uniformly at random. Note that in each row, the same operator is applied to the first qubit regardless of the column index. Similarly for the second qubit in each column. This means that Alice and Bob can perform the encoding of their own input independently of the input of the other participant. By translating these operators into the resulting state when applied to $|\varphi_{00}\rangle$, we get the states in Table 2, where one should note that—ignoring global phase factors—each of the "zero states" $|\varphi_{00}\rangle$, $|\varphi_{10}\rangle$, and $|\varphi_{01}\rangle$ correspond to inputs (0,0), (1,0), and (0,1) with equal probability when the encodings are chosen uniformly. The end effect is that Charlie receives the state $|\varphi_{11}\rangle$ if and only if $a = 1$ and $b = 1$, which is equivalent to $ab = 1$. If $(a, b) \neq (1, 1)$, Charlie will receive $|\varphi_{00}\rangle$, $|\varphi_{01}\rangle$, or $|\varphi_{10}\rangle$ with equal probability. That is, the specific encoding of zero received by Charlie reveals nothing about the individual inputs of Alice and Bob.[2]

Actually, the encodings that we have presented in this section do not quite match those that we propose in the general setting. More precisely, the three encodings in Table 1 are a subset of the encodings in the general procedure. The binary case has some extra symmetry compared to larger fields, and this allows a smaller family of encodings (three instead of six).

[2]Clearly, one can ask whether a similar type of operation is possible in a classical setting. In principle, it is, if Charlie has only access to the product without having access to the individual contributions; nevertheless, in a setting with entanglement this setup occurs naturally, without having any transmission toward Charlie, which is a genuine use of the "spooky action at a distance."

## IV. GENERAL CASE

In order to analyze products in $\mathbb{F}_p$ for general primes $p$, we will have a closer look at the operators applied by Alice and Bob. We will assume that Alice and Bob both use operators on the form $X(i)Z(j)$ where $i, j \in \mathbb{F}_p$. More precisely, for each input $a \in \mathbb{F}_p$ Alice will have values $x_a^A$ and $z_a^A$, both in $\mathbb{F}_p$, defining the operator $X(x_a^A)Z(z_a^A)$ that she will apply to her qubit. Note here that the superscript indicates that these values belong to Alice, and the subscript denotes the specific input. For instance, if $p = 2$ Alice's operators will be defined by values $x_0^A, z_0^A$ corresponding to $a = 0$ and $x_1^A, z_1^A$ corresponding to $a = 1$. In a similar way, we can define values $x_b^B$ and $z_b^B$ for Bob.

With this notation in place, the input $(a, b)$ will result in the quantum state

$$\left(X(x_a^A)Z(z_a^A) \otimes X(x_b^B)Z(z_b^B)\right) |\varphi_{00}\rangle. \tag{3}$$

There are, however, many different choices of $x_a^A, z_a^A, x_b^B$, and $z_b^B$ that lead to the same state. To handle this, we will make extensive use of the equivalence given in the following lemma.

*Lemma 1:* For any $x_a^A, z_a^A, x_b^B, z_b^B \in \mathbb{F}_p$ and any $(i, j) \in \mathbb{F}_p^2$, we have

$$\begin{aligned}\left(X(x_a^A)Z(z_a^A) \otimes X(x_b^B)Z(z_b^B)\right) |\varphi_{ij}\rangle \\ = \left(X(x_a^A - x_b^B)Z(z_a^A + z_b^B) \otimes I\right) |\varphi_{ij}\rangle\end{aligned}$$

up to a global phase.

*Proof:* Direct calculations reveal that

$$\begin{aligned}&\left(X(x_a^A)Z(z_a^A) \otimes X(x_b^B)Z(z_b^B)\right) |\varphi_{ij}\rangle \\ &= \omega^{i z_a^A} \sum_{k=0}^{p-1} \omega^{k(j+z_a^A+z_b^B)} |k + i + x_a^A\rangle |k + x_b^B\rangle\end{aligned}$$

and

$$\begin{aligned}&\left(X(x_a^A - x_b^B)Z(z_a^A + z_b^B) \otimes I\right) |\varphi_{ij}\rangle \\ &= \omega^{i(z_a^A+z_b^B)} \sum_{k=0}^{p-1} \omega^{k(j+z_a^A+z_b^B)} |k + i + x_a^A - x_b^B\rangle |k\rangle \\ &= \omega^{(i+x_b^B)(z_a^A+z_b^B)+z_b^B j} \sum_{k=0}^{p-1} \omega^{k(j+z_a^A+z_b^B)} |k+i+x_a^A\rangle |k+x_b^B\rangle\end{aligned}$$

where the last equality follows from appropriate substitution of the summing variable. ■

Lemma 1 not only gives us an equivalence between different operators, it also allows us to describe the Bell states using elements of $\mathbb{F}_p^2$. Namely, the state (3) can be uniquely represented by the pair $(x_a^A - x_b^B, z_a^A + z_b^B) \in \mathbb{F}_p^2$. We use this to describe the different encodings as was done in Section III.

Our strategy is to use this insight to describe "multiplication tables," i.e., to define $p \times p$ tables such that the Bell-like state in entry $(i, j)$ represents the product $ij$. Notationally, such a table corresponds to a bijection, and we will refer to this as an *encoding*.

*Definition 1:* An encoding is a bijection $\varepsilon : \mathbb{F}_p^2 \to \mathbb{F}_p^2$. The set of all encodings is denoted by $\mathcal{E}$.

Not all encodings match the properties needed to compute products, however, so we derive necessary and sufficient conditions for an encoding to be valid. First of all, the encoded Bell state must correspond to the correct product. In other words, if Alice and Bob have inputs $(a, b)$, and this is encoded to a Bell-state $|\varphi_{ij}\rangle$, then it must be the case that Charlie recognizes this as representing the product $ab$. That is, $ab = ij$ (where computations are done in $\mathbb{F}_p$).

A second condition comes from the way Alice and Bob apply the encodings to their qubits. More precisely, an encoding can only be used if there exist local operations represented by $\{(x_i^A, z_i^A)\}_{i\in\mathbb{F}_p}$ and $\{(x_i^B, z_i^B)\}_{i\in\mathbb{F}_p}$ that realize said encoding. It turns out that this is equivalent to $\varepsilon \in \mathcal{E}$ satisfying

$$\varepsilon(i, j) + \varepsilon(i', j') = \varepsilon(i, j') + \varepsilon(i', j)$$

for every $i, i', j, j'$. The analysis leading to this is somewhat involved, so we give it in Appendix A.

Summarizing this in a single definition, we get the encodings that we need. Here, we use

$$\pi : \begin{cases} \mathbb{F}_p^2 \to \mathbb{F}_p \\ (i, j) \mapsto ij \end{cases} \tag{4}$$

as a shorthand notation for products, which will make the exposition less cumbersome.

*Definition 2:* An encoding $\varepsilon \in \mathcal{E}$ is called product-compatible if it satisfies the following.

1) For every $i, j \in \mathbb{F}_p$, we have $\pi(\varepsilon(i, j)) = ij$.
2) For every $i, j, i', j' \in \mathbb{F}_p$ it holds that

$$\varepsilon(i, j) + \varepsilon(i', j') = \varepsilon(i, j') + \varepsilon(i', j).$$

*Definition 3:* A family of encodings $E \subseteq \mathcal{E}$ is called a private product family, if each $\varepsilon \in E$ satisfies the following.

1) $\varepsilon$ is product-compatible.
2) For every $(a, b) \in \mathbb{F}_p^2$, there exists $c \in \mathbb{Z}^+$ such that

$$\left|\{\varepsilon \in E \mid \varepsilon(i, j) = (a, b)\}\right| = c$$

for every $(i, j) \in \mathbb{F}_p^2$ satisfying $ij = ab$.

A few comments about the intuition behind Definition 3 are in order. The first condition is related to correctness, meaning that the state that Charlie receives will actually correspond to the correct product. The second requirement ensures privacy. It guarantees that when Charlie receives a state $|\varphi_{ab}\rangle$, then it will have come from *any* input pair $(i, j)$ satisfying $ij = ab$ with equal probability as long as Alice and Bob choose $\varepsilon \in E$ uniformly at random.

The proposed method for computing private products using entangled pairs can be found in Fig. 1.

1) Alice samples $\varepsilon \in E$ uniformly at random, where $E$ is a private product family. She sends a description of $\varepsilon$ to Bob via a classical channel.
2) Alice and Bob apply local operations to their qubit according to $\varepsilon$.
3) Alice and Bob both send (or teleport) their qubit to Charlie, who measures it in the basis formed by all states on the form (2). If he measures $|\varphi_{ij}\rangle$, the product is $ij = ab$.

**FIGURE 1. Protocol for entanglement-assisted private products.**

## V. CONSTRUCTING A PRIVATE PRODUCT FAMILY

In the following, we describe an explicit way to produce a private product family over arbitrary finite fields. In greater detail, we focus on a subset of $\mathcal{E}$ and obtain the private product family as orbits under group actions defined later. Our starting point will be two "canonical" product-compatible encodings given by $\varepsilon_0(i, j) = (i, j)$ and $\varepsilon_0^T(i, j) = (j, i)$. Intuitively, the idea is that these are "good" encodings in the sense that they posses the properties we want. By choosing the group actions appropriately (i.e., in a way that preserves the desired properties) we obtain additional "good" encodings by considering the orbits of $\varepsilon_0$ and $\varepsilon_0^T$.

In our proofs, we will rely on an additional property of $\varepsilon_0$ and $\varepsilon_0^T$ that is also preserved by the group action. Namely, we define $E_1 \subseteq \mathcal{E}$ by

$$E_1 = \left\{ \varepsilon \;\middle|\; \begin{array}{l} \varepsilon \text{ is product-compatible,} \\ \forall (i, j) \in \mathbb{F}_p^2 \ \forall \delta \in \mathbb{F}_p : \\ \varepsilon(i+\delta, 0) - \varepsilon(i, 0) = \varepsilon(j+\delta, 0) - \varepsilon(j, 0) \end{array} \right\}. \tag{5}$$

*Proposition 1:* We have $\varepsilon_0 \in E_1$ and $\varepsilon_0^T \in E_1$.

*Proof:* From the definition of $\varepsilon_0$ and $\varepsilon_0^T$, it is clear that they are product-compatible. In addition, we see that for any $(i, j) \in \mathbb{F}_p^2$ and $\delta \in \mathbb{F}_p$, $\varepsilon_0$ satisfies

$$\begin{aligned} \varepsilon_0(i+\delta, 0) - \varepsilon_0(i, 0) &= (i+\delta, 0) - (i, 0) \\ &= (i+\delta, 0) - (i, 0) + ((j-i, 0) - (j-i, 0)) \\ &= (j+\delta, 0) - (j, 0) \\ &= \varepsilon_0(j+\delta, 0) - \varepsilon_0(j, 0) \end{aligned}$$

as required. The proof for $\varepsilon_0^T$ is similar. ∎

Now, fix a primitive element $\alpha$ of $\mathbb{F}_p$ (i.e. $\alpha$ has multiplicative order $p-1$), and consider the additive group $\mathbb{Z}_{p-1} = \mathbb{Z}/(p-1)\mathbb{Z}$. For each $n \in \mathbb{Z}_{p-1}$ and each $\beta \in \mathbb{F}_p$, define a map $\varphi_{n,\beta} : E_1 \to E_1$ given by

$$\varphi_{n,\beta}(\varepsilon)(i, j) = \begin{cases} \varepsilon(\alpha^n i + \beta, 0) & j = 0 \\ \varepsilon(\alpha^n i, \alpha^{-n} j) & j \neq 0. \end{cases}$$

*Proposition 2:* For every $n \in \mathbb{Z}_{p-1}$ and $\beta \in \mathbb{F}_p$, the map $\varphi_{n,\beta} : E_1 \to E_1$ is well-defined.

*Proof:* Let $n \in \mathbb{Z}_{p-1}$, and assume $\varepsilon \in E_1$. We first show that $\varphi_{n,\beta}(\varepsilon)$ is product-compatible. Indeed, for $j = 0$ it is easy to check that $\pi(\varphi_{n,\beta}(\varepsilon)(i, j)) = 0 = ij$, and for $j \neq 0$ we have

$$\pi\left(\varphi_{n,\beta}(\varepsilon)(i, j)\right) = \pi\left(\varepsilon(\alpha^n i, \alpha^{-n} j)\right) = \alpha^n i \alpha^{-n} j = ij.$$

To see that condition 2 in Definition 2 is satisfied, notice that for $j \neq 0$, $j' \neq 0$, we have

$$\begin{aligned} &\varphi_{n,\beta}(\varepsilon)(i, j) + \varphi_{n,\beta}(\varepsilon)(i', j') \\ &\quad = \varepsilon(\alpha^n i, \alpha^{-n} j) + \varepsilon(\alpha^n i', \alpha^{-n} j') \\ &\quad = \varepsilon(\alpha^n i, \alpha^{-n} j') + \varepsilon(\alpha^n i', \alpha^{-n} j) \\ &\quad = \varphi_{n,\beta}(\varepsilon)(i, j') + \varphi_{n,\beta}(\varepsilon)(i', j) \end{aligned}$$

where the second equality stems from $\varepsilon$ being product compatible. Otherwise, we can by symmetry assume $j = 0$ and $j' \neq 0$, which implies

$$\begin{aligned} &\varphi_{n,\beta}(\varepsilon)(i, j) + \varphi_{n,\beta}(\varepsilon)(i', j') \\ &\quad = \varepsilon(\alpha^n i + \beta, 0) + \varepsilon(\alpha^n i', \alpha^{-n} j') \\ &\quad = \varepsilon(\alpha^n i + \beta, \alpha^{-n} j') + \varepsilon(\alpha^n i', 0) \\ &\quad = \varepsilon(\alpha^n i, \alpha^{-n} j') + \varepsilon(\alpha^n i' + \beta, 0) \\ &\quad = \varphi_{n,\beta}(\varepsilon)(i, j') + \varphi_{n,\beta}(\varepsilon)(i', j) \end{aligned}$$

where the first and second conditions in (5) give the second and third equalities, respectively.

For the remaining condition in (5), observe that for every $(i, j) \in \mathbb{F}_p^2$, and every $\delta \in \mathbb{F}_p$, we have

$$\begin{aligned} &\varphi_{n,\beta}(\varepsilon)(i+\delta, 0) - \varphi_{n,\beta}(\varepsilon)(i, 0) \\ &\quad = \varepsilon\left((\alpha^n i + \beta) + \alpha^n \delta, 0\right) - \varepsilon(\alpha^n i + \beta, 0) \\ &\quad = \varepsilon\left((\alpha^n j + \beta) + \alpha^n \delta, 0\right) - \varepsilon(\alpha^n j + \beta, 0) \\ &\quad = \varphi_{n,\beta}(\varepsilon)(j+\delta, 0) - \varphi_{n,\beta}(\varepsilon)(j, 0) \end{aligned}$$

where the second equality once again stems from $\varepsilon \in E_1$. Thus, $\varphi_{n,\beta}(\varepsilon) \in E_1$, and $\varphi_{n,\beta}$ is well-defined. ∎

Consider the additive group $\mathbb{Z}_{p-1}$ and $\mathbb{F}_p$ as a group with addition as well. By defining the composition $\circ : \mathbb{Z}_{p-1} \times \mathbb{F}_p \to \mathbb{Z}_{p-1} \times \mathbb{F}_p$, given by

$$(n, \beta) \circ (n', \beta') = (n + n', \alpha^{n'} \beta + \beta') \tag{6}$$

we obtain the semidirect product $G_1 = \mathbb{Z}_{p-1} \ltimes \mathbb{F}_p$. In more detail, the map $\beta \mapsto \alpha^{n'} \beta$ is an automorphism of $\mathbb{F}_p$ for any $n' \in \mathbb{Z}_{p-1}$, and the construction above yields the outer semidirect product as described in [6, p. 76].

*Proposition 3:* Let $G_1 = \mathbb{Z}_{p-1} \ltimes \mathbb{F}_p$ with composition as in (6). The map given by

$$\begin{cases} G_1 \times E_1 \to E_1 \\ ((n, \beta), \varepsilon) \mapsto \varphi_{n,\beta}(\varepsilon) \end{cases}$$

defines a group action of $G_1$ on $E_1$.

*Proof:* By Proposition 2, the map is well-defined. The following observations imply that it defines a group action. First, for every $\varepsilon \in E_1$, we have $\varphi_{0,0}(\varepsilon) = \varepsilon$. Second, we see

that

$$\begin{aligned}\varphi_{n,\beta}(\varphi_{n',\beta'}(\varepsilon))(i,0) &= \varphi_{n',\beta'}(\varepsilon)(\alpha^n i+\beta,0)\\ &= \varepsilon(\alpha^{n+n'}i+\alpha^{n'}\beta+\beta',0)\\ &= \varphi_{(n,\beta)\circ(n',\beta')}(\varepsilon)\end{aligned}$$

and

$$\begin{aligned}\varphi_{n,\beta}(\varphi_{n',\beta'}(\varepsilon))(i,j) &= \varphi_{n',\beta'}(\varepsilon)(\alpha^n i,\alpha^{-n}j)\\ &= \varepsilon(\alpha^{n+n'}i,\alpha^{-(n+n')}j)\\ &= \varphi_{(n,\beta)\circ(n',\beta')}(\varepsilon)(i,j)\end{aligned}$$

for $j\neq 0$. ∎

Having established that $G_1$ does indeed act on $E_1$, we verify that $\varepsilon_0$ and $\varepsilon_0^T$ are in different orbits and hence give rise to different encodings.

*Proposition 4:* The orbits $G_1\varepsilon_0$ and $G_1\varepsilon_0^T$ are disjoint. In addition, they satisfy $|G_1\varepsilon_0| = |G_1\varepsilon_0^T| = |G_1|$.

*Proof:* Note, that for any $(n,\beta)\in G_1$, we have $\varphi_{n,\beta}(\varepsilon_0)(0,1) = (0,\alpha^{-n})$, and similarly for any $(n',\beta')\in G_1$ it holds that $\varphi_{n',\beta'}(\varepsilon_0^T)(0,1) = (\alpha^{-n'},0)$. Hence, the two orbits are disjoint.

We show the cardinality claim for $G_1\varepsilon_0$ only, as $G_1\varepsilon_0^T$ is similar. Assume that $(n,\beta)$ and $(n',\beta')$ are two elements of $G_1$ that map $\varepsilon_0$ to the same element of $E_1$. In particular, this implies for $j\neq 0$ that $\varphi_{n,\beta}(\varepsilon_0)(i,j) = \varphi_{n',\beta'}(\varepsilon_0)(i,j)$, which implies $\alpha^{-n}j = \alpha^{-n'}j$. Since $j$ is assumed to be a unit, we conclude $n = n'$. Applying similar arguments to indices $(i,0)$ gives $\alpha^n i+\beta = \alpha^n i+\beta'$, implying $\beta=\beta'$. Thus, $G_1$ acts injectively on $E_1$. ∎

The second group action that we are going to use is very similar to $G_1$ acting on $E_1$. In fact, the group will be the same, but the action is different. Hence, we will denote the group by $G_2$ in connection to this new group action to ease the notation. This group $G_2$ will act on a second subset of $\mathcal{E}$ given by

$$E_2 = \left\{\varepsilon \;\middle|\; \begin{array}{l}\varepsilon \text{ is product-compatible,}\\ \forall (i,j)\in\mathbb{F}_p^2\ \forall\delta\in\mathbb{F}_p:\\ \varepsilon(0,i+\delta)-\varepsilon(0,i)=\varepsilon(0,j+\delta)-\varepsilon(0,j)\end{array}\right\}.$$

Each element $(n,\beta)\in G_2 = \mathbb{Z}_{p-1}\rtimes\mathbb{F}_p$ gives rise to a map $\psi_{n\beta}: E_2\to E_2$ given by

$$\psi_{n,\beta}(\varepsilon)(i,j) = \begin{cases}\varepsilon(0,\alpha^{-n}j+\beta) & i=0\\ \varepsilon(\alpha^n i,\alpha^{-n}j) & i\neq 0.\end{cases}$$

One can then prove that this is indeed a group action as was done in Propositions 2 and 3 for $G_1$, and analyze the orbits as in Proposition 4. For $G_2$ we simply state the results and omit the proofs, as they are completely analogous to the previous ones for $G_1$.

*Proposition 5:* Let $G_2 = \mathbb{Z}_{p-1}\rtimes\mathbb{F}_p$ with composition as in (6). The map given by

$$\begin{cases} G_2\times E_2 \to E_2\\ ((n,\beta),\varepsilon)\mapsto \psi_{n,\beta}(\varepsilon)\end{cases}$$

defines a group action of $G_2$ on $E_2$.

*Proposition 6:* The orbits $G_2\varepsilon_0$ and $G_2\varepsilon_0^T$ are disjoint. In addition, they satisfy $|G_2\varepsilon_0| = |G_2\varepsilon_0^T| = |G_2|$.

Now, define the sets

$$E = H_1\cup H_2,\quad H_i = G_i\varepsilon_0\cup G_i\varepsilon_0^T \tag{7}$$

which are all subsets of $\mathcal{E}$. The set $E$ is, we claim, a private product family as desired, and the following lemma and proposition prove this claim.

*Lemma 2:* Writing $E$ as a disjoint union

$$E = (H_1\setminus H_2)\sqcup(H_2\setminus H_1)\sqcup(H_1\cap H_2)$$

we have

$$\begin{aligned}H_1\setminus H_2 &= \{\varphi_{n,\beta}(\varepsilon)\mid \varepsilon\in\{\varepsilon_0,\varepsilon_0^T\},(n,\beta)\in G_1,\beta\neq 0\}\\ H_2\setminus H_1 &= \{\psi_{n,\beta}(\varepsilon)\mid \varepsilon\in\{\varepsilon_0,\varepsilon_0^T\},(n,\beta)\in G_2,\beta\neq 0\}\end{aligned}$$

and

$$\begin{aligned}H_1\cap H_2 &= \{\varphi_{n,0}(\varepsilon)\mid \varepsilon\in\{\varepsilon_0,\varepsilon_0^T\},(n,0)\in G_1\}\\ &= \{\psi_{n,0}(\varepsilon)\mid \varepsilon\in\{\varepsilon_0,\varepsilon_0^T\},(n,0)\in G_2\}.\end{aligned}$$

In addition, $|H_1\setminus H_2| = |H_2\setminus H_1| = 2(p-1)^2$ and $|H_1\cap H_2| = 2(p-1)$.

*Proof:* It is clear that the union is disjoint by definition.

Assuming that $h\in H_1\cap H_2$, there must exist $n,n'\in\mathbb{Z}_{p-1}$, $\beta,\beta'\in\mathbb{F}_p$, and $\varepsilon,\varepsilon'\in\{\varepsilon_0,\varepsilon_0^T\}$ such that $\varphi_{n,\beta}(\varepsilon) = h = \psi_{n',\beta'}(\varepsilon')$. Observe first that

$$\begin{aligned}\varphi_{n,\beta}(\varepsilon)(0,0) &= \varepsilon(\beta,0)\\ \psi_{n',\beta'}(\varepsilon')(0,0) &= \varepsilon'(0,\beta').\end{aligned}$$

So for $\varphi_{n,\beta}(\varepsilon) = \psi_{n',\beta'}(\varepsilon')$ to hold, it must be the case that either $\beta=\beta'=0$, or $\beta=\beta'\neq 0$ and $\varepsilon\neq\varepsilon'$. The latter case is impossible, however, as

$$\begin{aligned}\varphi_{n,\beta}(\varepsilon)(0,1) &= \varepsilon(0,\alpha^n)\\ \psi_{n',\beta}(\varepsilon')(0,1) &= \varepsilon'(0,\alpha^{n'}+\beta)\end{aligned}$$

and regardless of the choices of $\varepsilon\neq\varepsilon'$, this implies $\alpha^n = 0$, which is a contradiction. Hence, $\beta=\beta'=0$. For these mappings, observe that they satisfy

$$\varphi_{n,0}(\varepsilon)(i,j) = \varepsilon(\alpha^n i,\alpha^{-n}j) = \psi_{n,0}(\varepsilon)(i,j)$$

regardless of the choice of $\varepsilon$ and the values of $i$ and $j$, so they constitute the elements of $H_1\cap H_2$ as claimed.

What remains is to prove that $|H_1\cap H_2| = 2(p-1)$, as the cardinalities of $H_1\setminus H_2$ and $H_2\setminus H_1$ then follow from Propositions 4 and 6, respectively. From Proposition 4, we know that $G_1$ acts injectively on $E_1$, meaning that each of the $\varphi_{n,0}(\varepsilon)$ appearing in the statement of the Lemma are distinct. The $p-1$ choices for $n \in \mathbb{Z}_{p-1}$ and the 2 choices for $\varepsilon \in \{\varepsilon_0, \varepsilon_0^T\}$ then give $|H_1 \cap H_2| = 2(p-1)$ as claimed. ∎

*Proposition 7:* The set $E$ defined in (7) is a private product family.

*Proof:* Since all the orbits making up $E$ consist of product-compatible encodings, we only need to prove item 2 in Definition 3. Fix $(a, b), (i, j) \in \mathbb{F}_p^2$ such that $ab = ij$, and assume first that $ab \neq 0$. Consider $\varphi_{n,\beta}(\varepsilon_0) \in (H_1 \setminus H_2)$ as given in Lemma 2. If $\varphi_{n,\beta}(\varepsilon_0)(i, j) = (a, b)$, then $\varepsilon_0(\alpha^n i, \alpha^{-n} j) = (a, b)$, which again implies $\alpha^n = ai^{-1}$. Here, we use that $i$ is a nonzero element of $\mathbb{F}_p$. Note also that the assumption $ab = ij$ means that $a = \alpha^n i$ automatically implies $b = \alpha^{-n} j$ as needed. Since $\alpha$ is primitive, there is exactly one $n$ that satisfies $\alpha^n = ai^{-1}$, and $\beta \in \mathbb{F}_p^*$ can be chosen freely. Thus, this gives $p-1$ encodings $\varepsilon \in H_1 \setminus H_2$ such that $\varepsilon(i, j) = (a, b)$, and similar arguments applied to $\varepsilon_0^T$ yields another $p-1$ encodings. Completely analogously, there are $2(p-1)$ such encodings in $H_2 \setminus H_1$. In the case of $H_1 \cap H_2$, there is only a single choice for $\alpha$ as above, and in addition $\beta = 0$ in this case. Hence, one would find 2 encodings in $H_1 \cap H_2$. In total, this amounts to $4(p-1) + 2$ encodings regardless of the choice of $(i, j)$.

Moving on, assume $ab = 0$ with $a \neq 0$ – the case $b \neq 0$ is analogous. Considering $(i, 0) \in \mathbb{F}_p$ with $i \neq 0$, the possible encodings are

$$(a, 0) = \varphi_{n,\beta}(\varepsilon_0)(i, 0) = (\alpha^n i + \beta, 0)$$
$$(a, 0) = \psi_{n,\beta}(\varepsilon_0)(i, 0) = (\alpha^n i, 0).$$

For the first equality, each choice of $n \in \mathbb{Z}_{p-1}$ gives a unique choice of $\beta$. In the second, there is one possible $n$, but $\beta \in \mathbb{F}_p$ can be chosen freely. Note, however, that there is an encoding $\varphi_{n,\beta}(\varepsilon_0) = \psi_{n,\beta}(\varepsilon_0)$ (as shown in Lemma 2) that is counted twice in this way. Hence, the total number of encodings $\varepsilon \in E$ satisfying $\varepsilon(i, 0) = (a, 0)$ is $(p-1+p)-1 = 2(p-1)$. The same strategy can be used to show that there are also $2(p-1)$ encodings such that $\varepsilon(0, j) = (a, 0)$ by considering $\varphi_{n,\beta}(\varepsilon_0^T)$ and $\psi_{n\beta}(\varepsilon_0^T)$. Thus, regardless of the choice of $(i, j)$ with $ij = ab = 0$, there are exactly $2(p-1)$ encodings $\varepsilon \in E$ satisfying $\varepsilon(i, j) = (a, b)$.

Finally, consider $(a, b) = (0, 0)$. The candidate encodings if $(i, j) = (0, 0)$ are

$$(0, 0) = \varphi_{n,\beta}(\varepsilon_0)(0, 0) \ = (\beta, 0)$$
$$(0, 0) = \varphi_{n,\beta}(\varepsilon_0^T)(0, 0) \ = (0, \beta)$$
$$(0, 0) = \psi_{n,\beta}(\varepsilon_0)(0, 0) \ = (0, \beta)$$
$$(0, 0) = \psi_{n,\beta}(\varepsilon_0^T)(0, 0) = (\beta, 0).$$

Observe that $\beta = 0$ in any case, so $\varphi_{n,\beta}(\varepsilon_0) = \psi_{n,\beta}(\varepsilon_0)$ and $\varphi_{n,\beta}(\varepsilon_0^T) = \psi_{n,\beta}(\varepsilon_0^T)$. For each, any $n$ is possible, giving $2(p-1)$ encodings. For $(i, j) = (i, 0)$ with $i \neq 0$, we use a similar strategy and consider

$$(0, 0) = \varphi_{n,\beta}(\varepsilon_0)(i, 0) \ = (\alpha^n i + \beta, 0)$$
$$(0, 0) = \varphi_{n,\beta}(\varepsilon_0^T)(i, 0) = (0, \alpha^n i + \beta).$$

For each $n \in \mathbb{Z}_{p-1}$ there is a unique choice of $\beta$, and all of these encodings are distinct by Proposition 4. Thus, we have $2(p-1)$ encodings like previously. If $(i, j) = (0, j)$, similar arguments can be applied to $\psi_{n,\beta}(\varepsilon_0)$ and $\psi_{n,\beta}(\varepsilon_0^T)$. ∎

## VI. SYSTEMATIC CHOICES

The analysis in Section IV provides a way to permute the Bell states similarly to what was done in Section III. But by considering these encodings carefully, it also gives us a systematic way to choose $x_a^A, z_a^A, x_b^B, z_b^B$ such that the operators applied by Alice and Bob correspond to using a specific encoding from $E$. Namely, to perform the encoding according to $\varphi_{n,\beta}(\varepsilon_0)$ Alice will set $x_a^A = \alpha^n a$ and $z_a^A = 0$. Bob will use $z_b^B = \alpha^{-n} b$ and

$$x_b^B = \begin{cases} -\beta & b = 0 \\ 0 & b \neq 0. \end{cases}$$

Using Lemma 1, this implies that Alice and Bob will end up in state

$$\big(X(\alpha^n a) \otimes X(-\beta)\big) |\varphi_{00}\rangle = \big(X(\alpha^n a + \beta) \otimes I\big) |\varphi_{00}\rangle$$

if $b = 0$, and

$$\big(X(\alpha^n a) \otimes Z(\alpha^{-n} b)\big) |\varphi_{00}\rangle = \big(X(\alpha^n a) Z(\alpha^{-n} b) \otimes I\big) |\varphi_{00}\rangle$$

otherwise. In any case, this corresponds exactly to the encoding $\varphi_{n,\beta}(\varepsilon_0)$, as

$$\varphi_{n,\beta}(\varepsilon_0)(a, b) = \begin{cases} (\alpha^n a + \beta, 0) & b = 0 \\ (\alpha^n a, \alpha^{-n} b) & b \neq 0. \end{cases}$$

Similar considerations can be done for the remaining encodings in $E$, leading to the systematic choices presented in Table 3.

We note that when applying this private product family in Fig. 1, one way to sample $\varepsilon \in E$ uniformly is to first sample a trit $T \in \{1, 2, 3\}$ with probabilities $\Pr[T = 1] = \Pr[T = 2] = (p-1)/(2p-1)$ and $\Pr[T = 3] = 1/(2p-1)$. Each outcome then corresponds to one of the cases in Lemma 2 with probabilities matching the proportion of $\varepsilon$ contained in each case. After that, Alice can simply sample $(n, \beta)$ according to the requirements in the case determined by $T$.

To illustrate the use of the protocol in Fig. 1, we provide two examples.

*Example 1:* Let $p = 5$, $\alpha = 2$, and assume that Alice and Bob have inputs $a = 2$ and $b = 4$, respectively. Alice samples an encoding $\varepsilon$ from $E$ uniformly at random, and for concreteness we use $\varepsilon = \psi_{3,2}(\varepsilon_0)$ in this example. She sends this outcome to Bob.

Following the systematic method in Table 3, Alice now applies the operator $X(2^3 \cdot 2)Z(0) = X(1)$ to her qubit, and Bob applies $X(0)Z(2^{-3} \cdot 4) = Z(3)$ to his. The overall state of the system is now

$$(X(1) \otimes Z(3)) |\varphi_{00}\rangle = |\varphi_{13}\rangle$$

where we again omit the overall phase. Thus, when they each send their qubit to Charlie, the state $|\varphi_{13}\rangle$ reveals that the product is $1 \cdot 3 = 3$, which is consistent with $ab = 2 \cdot 4 = 3$.

**TABLE 3.** Systematic Choices for Parameters $x_a^A, z_a^A, x_b^B, z_b^B$ When Using the Private Product Family $E$ Defined in (7)

| | $\varphi_{n,\beta}$ | $\psi_{n,\beta}$ |
|---|---|---|
| $\varepsilon_0$ | $x_a^A = \alpha^n a,\ z_b^B = \alpha^{-n} b,\ z_a^A = 0$ <br> $x_b^B = \begin{cases} -\beta & b = 0 \\ 0 & b \neq 0. \end{cases}$ | $x_a^A = \alpha^n a,\ z_b^B = \alpha^{-n} b,\ x_b^B = 0$ <br> $z_a^A = \begin{cases} \beta & a = 0 \\ 0 & a \neq 0. \end{cases}$ |
| $\varepsilon_0^T$ | $z_a^A = \alpha^n a,\ x_b^B = \alpha^{-n} b,\ x_a^A = 0,$ <br> $z_b^B = \begin{cases} \beta & b = 0 \\ 0 & b \neq 0. \end{cases}$ | $z_a^A = \alpha^n a,\ x_b^B = \alpha^{-n} b,\ z_b^B = 0$ <br> $x_a^A = \begin{cases} \beta & a = 0 \\ 0 & a \neq 0. \end{cases}$ |

Note also, that $\psi_{3,2}(\varepsilon_0)(a, b) = (2^3 a, 2^{-3} b) = (1, 3)$ which is in agreement with the operators applied by Alice and Bob since $X(1)Z(3) \otimes I$ and $X(1) \otimes Z(3)$ are equivalent w.r.t. the Bell-like states according to Lemma 1.

*Example 2:* Consider the same situation as in Example 1, but with $a = 0$. Assuming that the same encoding, $\psi_{3,2}(\varepsilon_0)$, is used, Alice will instead apply $X(2^3 \cdot 0)Z(2) = Z(2)$, while Bob applies $Z(3)$ as before. The overall state is then

$$(Z(2) \otimes Z(3)) |\varphi_{00}\rangle = (I \otimes I) |\varphi_{00}\rangle = |\varphi_{00}\rangle$$

and Charlie infers the product $0 \cdot 0 = 0$ as expected. Again, one may note that $\psi_{3,2}(\varepsilon_0)(0, 4) = (0, 2^{-3} b + 2) = (0, 0)$, describing the same state $|\varphi_{00}\rangle$ that Alice and Bob constructed above.

## VII. EXTENSION TO DOT PRODUCTS

In the binary case, the protocol in Fig. 1 can be easily extended to compute a PSI or a private dot product. Extending it to PSI is the easiest, as each possible set element $e_i$ receives an index $i$, and Alice and Bob then set $a_i = 1$ and $b_i = 1$, respectively, if $e_i$ is contained in their individual sets. Applying the protocol in a component-wise fashion then reveals exactly the set intersection to Charlie. Note, however, that this is not a scalable approach, as the required number of products is given by the size of the set domain. Thus, this approach is only feasible for smaller examples and more advanced techniques must be used in general, see e.g., [7], [8], [9], [10].

Altering this to a private dot product only requires Alice to sample a uniformly random permutation of the indices, inform Bob of the outcome, and then have them both apply this permutation to the ordering of the Bell states before sending them to Charlie. In this way, Charlie only learns the number of indices $i$ such that $a_i = 1$ and $b_i = 1$. But this is exactly the same as the dot product.

## VIII. CONCLUSION AND OPEN PROBLEMS

In this article, we showed that private products over finite fields can be computed by sacrificing a pair of entangled qubits. Moreover, the set defined in (7) provides an explicit description of encodings that allow this computation to happen over fields $\mathbb{F}_p$ for arbitrary choice of prime $p$.

The idea presented here could be extended in several ways. First, one could analyze if a similar approach is possible for general finite fields, $\mathbb{F}_q$ with $q = p^r$ a power of a prime. Another direction is to consider more than two inputting parties, thus aiming to compute the product of $n$ inputs while still giving the output to a participants with no input (like Charlie in the current article).

## APPENDIX A
## CONDITION FOR PRODUCT COMPATIBILITY

Let $\varepsilon \in \mathcal{E}$ be an encoding, and let $\varepsilon(i, j) = (\alpha_{ij}, \beta_{ij})$, meaning that we fix $\alpha_{ij}$ and $\beta_{ij}$ and want to find $(x_i^A, z_i^A)$ and $(x_j^B, z_j^B)$. In order for Alice and Bob to arrive at this state using local operations, it must by Lemma 1 be the case that their $(x_i^A, z_i^A)$ and $(x_j^B, z_j^B)$ are solutions to the linear system

$$\begin{cases} x_i^A - x_j^B = \alpha_{ij} \\ z_i^A + z_j^B = \beta_{ij}. \end{cases} \tag{8}$$

One would find such a system for every possible pair $(i, j)$. The $x_a^A, z_a^A, x_b^B$, and $z_b^B$ must be solutions to *all* of these systems simultaneously, meaning that we obtain $2p^2$ equations in $4p$ unknowns. One may note, however, that the $x$-part can be solved separately from the $z$-part, which instead gives two systems of $p^2$ equations in $2p$ unknowns. Considering the system concerning the $z$'s, it can be represented in matrix-form as given in (10), where the horizontal lines separate $p \times 2p$-matrices.

The system describing the $x$'s is similar, but with the $p$ last columns of the coefficient matrix multiplied by $p$-1 [caused by the change of sign in (8)]. Rather than working with this system as the linear combination of $p^2$-dimensional column vectors, we will consider equivalent $p \times p$-matrices. Namely, we define for each $i \in \{0, 1, \ldots, p-1\}$ the $p \times p$ matrices $R_i$ and $C_i$ with entries

$$(R_i)_{st} = \begin{cases} 1 & s \equiv i \pmod p \\ 0 & \text{otherwise} \end{cases}$$

and

$$(C_i)_{st} = \begin{cases} 1 & t \equiv i \pmod p \\ 0 & \text{otherwise,} \end{cases}$$

where row and column indexing start from 0. With this definition, the system in (10) can be represented as

$$\sum_{i=0}^{p-1} z_i^A R_i + \sum_{i=0}^{p-1} z_i^B C_i = M_\beta \tag{9}$$

$$\left[\begin{array}{cccccccccc} 1 & 0 & 0 & \cdots & 0 & 0 & 1 & 0 & \cdots & 0 \\ 1 & 0 & 0 & \cdots & 0 & 0 & 0 & 1 & \cdots & 0 \\ \vdots & \vdots & \vdots & \ddots & \vdots & \vdots & \vdots & \vdots & \ddots & \vdots \\ 1 & 0 & 0 & \cdots & 0 & 0 & 0 & 0 & \cdots & 1 \\ \hline 0 & 1 & 0 & \cdots & 0 & 0 & 1 & 0 & \cdots & 0 \\ 0 & 1 & 0 & \cdots & 0 & 0 & 0 & 1 & \cdots & 0 \\ \vdots & \vdots & \vdots & \ddots & \vdots & \vdots & \vdots & \vdots & \ddots & \vdots \\ 0 & 1 & 0 & \cdots & 0 & 0 & 0 & 0 & \cdots & 1 \\ \hline \vdots & \vdots & \vdots & \ddots & \vdots & \vdots & \vdots & \vdots & \ddots & \vdots \\ \hline 0 & 0 & 0 & \cdots & 0 & 1 & 1 & 0 & \cdots & 0 \\ 0 & 0 & 0 & \cdots & 0 & 1 & 0 & 1 & \cdots & 0 \\ \vdots & \vdots & \vdots & \ddots & \vdots & \vdots & \vdots & \vdots & \ddots & \vdots \\ 0 & 0 & 0 & \cdots & 0 & 1 & 0 & 0 & \cdots & 1 \end{array}\right] \begin{bmatrix} z_0^A \\ z_1^A \\ \vdots \\ z_{p-1}^A \\ z_0^B \\ z_1^B \\ \vdots \\ z_{p-1}^B \end{bmatrix} = \begin{bmatrix} \beta_{00} \\ \beta_{01} \\ \vdots \\ \beta_{0(p-1)} \\ \beta_{10} \\ \beta_{11} \\ \vdots \\ \beta_{1(p-1)} \\ \vdots \\ \beta_{(p-1)0} \\ \beta_{(p-1)1} \\ \vdots \\ \beta_{(p-1)(p-1)} \end{bmatrix} \tag{10}$$

where $M_\beta$ has entries $(M_\beta)_{ij} = \beta_{ij}$. Using this representation will simplify our arguments below, but before stating the result, we illustrate the notation in an example.

*Example 3:* If $p = 3$, the system in (10), shown at the top of this page, is

$$\begin{bmatrix} 1 & 0 & 0 & 1 & 0 & 0 \\ 1 & 0 & 0 & 0 & 1 & 0 \\ 1 & 0 & 0 & 0 & 0 & 1 \\ 0 & 1 & 0 & 1 & 0 & 0 \\ 0 & 1 & 0 & 0 & 1 & 0 \\ 0 & 1 & 0 & 0 & 0 & 1 \\ 0 & 0 & 1 & 1 & 0 & 0 \\ 0 & 0 & 1 & 0 & 1 & 0 \\ 0 & 0 & 1 & 0 & 0 & 1 \end{bmatrix} \begin{bmatrix} z_0^A \\ z_1^A \\ z_2^A \\ z_0^B \\ z_1^B \\ z_2^B \end{bmatrix} = \begin{bmatrix} \beta_{00} \\ \beta_{01} \\ \beta_{02} \\ \beta_{10} \\ \beta_{11} \\ \beta_{12} \\ \beta_{20} \\ \beta_{21} \\ \beta_{22} \end{bmatrix}.$$

In addition, columns 0 and 4 in the coefficient matrix correspond exactly to the $3 \times 3$-matrices

$$R_0 = \begin{bmatrix} 1 & 1 & 1 \\ 0 & 0 & 0 \\ 0 & 0 & 0 \end{bmatrix} \qquad C_1 = \begin{bmatrix} 0 & 1 & 0 \\ 0 & 1 & 0 \\ 0 & 1 & 0 \end{bmatrix}$$

when ordering the entries in a row-wise fashion.

*Definition 4:* Let $M$ be a $p \times p$-matrix over $\mathbb{F}_p$. We say that $M$ has property $\mathcal{P}$ if for every $(i, j) \in \mathbb{F}_p$ and $(i', j') \in \mathbb{F}_p$ it holds that

$$m_{ij} + m_{i'j'} = m_{ij'} + m_{i'j}$$

where computations are done in $\mathbb{F}_p$.

*Proposition 8:* Let $M_\beta$ be a $p \times p$-matrix over $\mathbb{F}_p$. Then, (9) has a solution if and only if $M_\beta$ has property $\mathcal{P}$ as defined in Definition 4.

*Proof:* Note first that if two matrices $A$ and $B$ satisfy $\mathcal{P}$, then $A + B$ satisfies $\mathcal{P}$ as well. In addition, it is easily checked that $R_i$ and $C_i$ satisfy $\mathcal{P}$ for every $i \in \{0, 1, \ldots, p-1\}$. This shows the "only if" part.

For the other direction, note that the number of $p \times p$-matrices satisfying $\mathcal{P}$ is $p^{2p-1}$. Namely, choosing the entries in the first row and column fixes all other entries. We show that this is exactly the number of matrices in the span of the $R_i$ and $C_i$ on the left-hand side of (9). The result then follows by the first part of the proof.

We claim that $\mathcal{B} = \{R_i\}_{i=0}^{p-1} \cup \{C_i\}_{i=1}^{p-1}$ is a basis for $\text{Span}(\{R_i\}_{i=0}^{p-1} \cup \{C_i\}_{i=0}^{p-1})$. To see this, note that all elements of $\{C_i\}_{i=1}^{p-1}$ has only zeros in column 0. Thus, the equation

$$\sum_{i=0}^{p-1} s_i R_i + \sum_{i=1}^{p-1} t_i C_i = O_{p \times p}$$

where $O_{p\times p}$ denotes the $p \times p$-dimensional zero matrix, implies that $s_i = 0$ for all $i$, and hence also $t_i = 0$ for all $i$. As such, $\mathcal{B}$ is linearly independent, and

$$C_0 = \sum_{i=0}^{p-1} R_i - \sum_{i=1}^{p-1} C_i$$

shows that $\text{Span}\,\mathcal{B} = \text{Span}(\{R_i\}_{i=0}^{p-1} \cup \{C_i\}_{i=0}^{p-1})$, meaning that $\mathcal{B}$ is a basis. Thus, there are $p^{2p-1}$ matrices in the span of $\{R_i\}_{i=0}^{p-1} \cup \{C_i\}_{i=0}^{p-1}$, concluding the proof. ■

*Remark 1:* The same result holds for the system describing the $x$-values. In particular, the only difference is a scalar on the $C_i$, but this does not change their span.

## ACKNOWLEDGMENT

The authors would like to thank the anonymous reviewers for pointing out aspects of our scheme that helped position it precisely.

## REFERENCES

[1] C. H. Bennett and S. J. Wiesner, "Communication via one- and two-particle operators on Einstein-Podolsky-Rosen states," *Phys. Rev. Lett.*, vol. 69, no. 20, 1992, Art. no. 2881, doi: 10.1103/PhysRevLett.69.2881.

[2] M. A. Nielsen and I. L. Chuang, *Quantum Computation and Quantum Information*, 10th ed. Cambridge, U.K.: Cambridge Univ. Press, 2010, doi: 10.1017/CBO9780511976667.

[3] A. Shamir, “How to share a secret,” *Commun. ACM*, vol. 22, no. 11, pp. 612–613, Nov. 1979, doi: 10.1145/359168.359176.

[4] R. Cramer, I. B. Damgård, and J. B. Nielsen, *Secure Multiparty Computation and Secret Sharing*. Cambridge, U.K.: Cambridge Univ. Press, 2015, doi: 10.1017/CBO9781107337756.

[5] A. Ketkar, A. Klappenecker, S. Kumar, and P. K. Sarvepalli, “Nonbinary stabilizer codes over finite fields,” *IEEE Trans. Inf. Theory*, vol. 52, no. 11, pp. 4892–4914, Nov. 2006, doi: 10.1109/TIT.2006.883612.

[6] S. Lang, *Algebra*, 3rd ed. ser. Graduate Texts in Mathematics Series, vol. 211, Berlin, Germany: Springer, 2002, doi: 10.1007/978-1-4613-0041-0.

[7] M. J. Freedman, K. Nissim, and B. Pinkas, “Efficient private matching and set intersection,” in *Advances in Cryptology - EUROCRYPT*, C. Cachin and J. L. Camenisch Eds. Berlin, Heidelberg: Springer, 2004, pp. 1–19, doi: 10.1007/978-3-540-24676-3_1.

[8] B. Pinkas, T. Schneider, and M. Zohner, “Scalable private set intersection based on OT extension,” *ACM Trans. Privacy Secur.*, vol. 21, no. 2, pp. 1–35, Jan. 2018, doi: 10.1145/3154794.

[9] M. Chase and P. Miao, “Private set intersection in the Internet setting from lightweight oblivious PRF,” in *Advances in Cryptology–CRYPTO*, D. Micciancio and T. Ristenpart Eds., Cham: Springer, 2020, pp. 34–63, doi: 10.1007/978-3-030-56877-1_2.

[10] B. Pinkas, M. Rosulek, N. Trieu, and A. Yanai, “PSI from paXos: Fast, malicious private set intersection,” in *Advances in Cryptology–EUROCRYPT*, 2020, A. Canteaut and Y. Ishai, Eds. Cham, Switzerland: Springer, pp. 739–767, doi: 10.1007/978-3-030-45724-2_25.